# New Approaches to Old Problems? Thinking About a New Design of the AML/CFT Strategy

Chiara Ferri*†

**Abstract.** The entry of new technological infrastructures into the financial markets poses serious concerns about the misuse of the economic system for illicit purposes, such as money laundering and financing of terrorism. Although there are cases in which this connection has already been discovered by malicious actors, distributed ledger technologies can nevertheless represent a powerful tool at the disposal of competent authorities to trace illicit flows and to better monitor risks in financial markets. However, this possibility may go through an interdisciplinary analysis of the phenomena. The search for alternative systems to move funds, rather than the traditional financial intermediaries, such as banks, is not a new circumstance and not necessarily for criminal purposes. Nevertheless, some of the already-known value transfer systems may benefit from the use of distributed ledger technology and make their detection more difficult. The European institutions are discussing the needed legislative packages to enforce the current regulations and to extend their application to the crypto space, as well as the establishment of a new competent authority.

KEYWORDS

1. Money Laundering.   2. Financing of Terrorism.   3. DLTs.
4. Financial Markets.   5. Financial Intelligence.   6. Blockchain analysis

## 1. Introduction

The possibility to transform and improve financial processes using emerging (and emerged) technological infrastructures, known as financial technology (or 'Fintech'), represents a new panorama of possibilities, but at the same time, also the concrete risk of the abuse of the system.

Moreover, authorities around the world are struggling with "decentralised finance" (DeFi) and the general role of distributed ledger technology (DLT). As well as affecting economic stability, the possibility to identify – and then use – the gap in current legislation provides opportunities for criminal activity.

Because of the increasing need for multidisciplinary teams consisting of together specialists from financial crime, organised crime, crypto and cyber units,[1] the need to reshape the current Anti-Money Laundering and Countering Financing of Terrorism regulation is one of the European Union's first commitments, not only as a legislative but also as a strategic activity, with the creation of a new competent authority and the strengthening of the existing intelligence network.

The aim of the proposals on the table of the European institutions is thus to improve the detection of suspicious transactions and activities and to close loopholes exploited by criminal to launder illicit proceeds or finance terrorist activities through the misuse of the financial system.[2]

---

†Chiara Ferri (chiara.ferri@unifi.it) is a PhD student in Blockchain & DLTs at the ISAS of the University of Camerino and member of BABEL, research unit of the Department of Economics and Management, University of Florence | Italy.

The new concept of 'embedded supervision', which comes from the idea that DLTs can be used to better monitor risks in financial markets, is, therefore, interesting.[3] Many approaches can be imagined, but at the basis, there must be a re-thinking of the strategy following two directions: the identification of the sources used by criminals and the prevention of the movement of funds, targeting the sources of funding. There is, therefore, a need to question how blockchain technology can be used and which are the new tools approaching in the financial sector.

This research is intended to be the starting point for a more comprehensive study; the aim is to identify the key cornerstones on which to focus if deemed as convincing. The paper will discuss the phenomena of money laundering and financing of terrorism as follows: in paragraph 2, it presents the 'state of the art' of the Anti-Money Laundering and Countering Financing of Terrorism (AML/CFT) framework within the European Union, underlying the differences between the two phenomena. It describes the strategy adopted by the European institution as well as the legislative proposals that are on the table. Paragraph 3 analyses the reflection of the money laundering and financing of terrorism in the economic sector to pinpoint why it is crucial to preserve the financial system from its misuse by criminal activities, with the example of the so-called Informal Value Transfer Systems (IVTS); the paragraph focuses on how the technology may represent a misuse of the abovementioned systems. Specifically, after examining the actual use of blockchains in the hawala system, the interest is pointed toward the possibility of its use to 'follow the money' to monitor the financial flows. Then, the prospective use of deposit tokens is briefly discussed. Paragraph 4 introduces the newly established Anti-Money Laundering Authority (AMLA) and its role within the existing Financial Intelligence Units (FIUs) as shaped by the proposed regulation, suggesting the need for a reshaping of the AML/CFT strategy based on the crucial role of technological innovation.

## 2. Anti-Money Laundering and Countering Financing of Terrorism: the 'state of the art'

The European strategy aimed to combat and prevent money laundering (ML) and the financing of terrorism (FT) dates back to 1990 with the first anti-money laundering directive; over the years, it has been modified several times due to multiple reasons, as the understanding of the need to address the evolving criminal tools and attitudes.

The first anti-money laundering (AML) directive was adopted in 1991.[4] It has been amended by each subsequent directive: in 2000,[5] 2005,[6] and 2015,[7] until the one currently in force, the fifth directive. Due to the increasing use of new technologies and to maintain the European legislative framework updated with the criminal trends, the European Union institutions are currently working on a new directive.[8]

Based on these considerations, on 18th January 2024, Parliament and Council negotiators reached a provisional agreement on the newly proposed AML/CFT package aimed to protect the EU's internal market from money laundering and terrorist financing.

The package, introduced in 2021, consists of a 6th Directive on AML/CFT ('AMLD6') to amend the fifth directive currently in force; the revision of Regulation 2015/847/EU on the traceability of crypto assets; a new regulation on AML/CFT ('AMLR') and the proposal for a regulation establishing a new European Union AML/CFT authority ('AMLA').

Although the two phenomena are often considered as part of the same issue, it is important to understand the differences between them and, specifically, that the financing of terrorism obtains the required liquidity through formal channels and the legal economy. Recently, one more phenomenon has been considered within the AML/CFT strategy, named the financing of the proliferation of weapons of mass destruction (WMD) or proliferation financing (PF).[9]

The fight against money laundering has a history dating back to 1980, with the first recommendation adopted by the Council in Strasbourg. In 1988, the Basel Committee on Banking Supervision adopted the General Statement of Ethical Principles on the Prevention of the Criminal Use of the Banking System for the Purpose of Money Laundering. The United Nations Convention against Illicit Traffic in Narcotic Drugs and Psychotropic Substances (1988), also known as the Vienna Convention, was the first inter-governmental agreement to criminalise money laundering.

Although the focus was on the increasing problems caused by the drug market, the signing states agreed to the obligation of criminalising the laundering of money derived from drug trafficking.

Even at its inception, money laundering was already known as a global phenomenon, leading to the establishment of the Financial Action Task Force (FATF) by the governments of the G-7 countries at their 1989 Economic Summit and currently one of the leading watchdogs in the AML/CFT field, with the primary responsibility for developing worldwide standards. Although there are several important milestones, it is possible to identify the first step in the modern AML/CFT era in June 1991 when the Council adopted the first directive on the 'prevention of the use of the financial system for the purpose of money laundering'.[10]

The term 'money laundering' has been addressed starting with the Vienna Convention and, at a definitional level, money laundering thus indicates that the modality through which the purchasing power of illegal origin becomes from potential to actual: the aim of anti-money laundering regulations is therefore to prevent the integration of illicitly obtained funds into the financial system, which can threaten its stability and legality.

Money laundering is defined as "the conversion or transfer of property, knowing that such property is derived from any offence(s), for the purpose of concealing or disguising the illicit origin of the property or of assisting any person who is involved in such offence(s) to evade the legal consequences of his actions".[11]

The focus is thus on the illegal origin of the liquidity. However, when it comes to the financing of terrorism, defined as the raising and processing of funds to supply terrorists with resources, it is important to consider that such activities are aimed at concealing the destination of the funds; it is a fact that generally, they have a legal origin.

Even though there are few differences in the two phenomena, the literature identifies similar stages: the collection, during which funds of both legal and illegal origin are gathered by a main collector; the transmission or dissimulation (through the use of 'underground', 'parallel' and alternative payment systems, rather than the conventional banking circuit) of the funds to conceal their ultimate purpose and, lastly, the use of the money or other assets.

Specifically for the financing of terrorism, it is possible to identify two pillars of the strategy, such as the 'freezing' measures, which involve the operational freezing of capital and economic resources to act directly on the sources and all the measures similar to anti-money laundering measures, which aim to hinder the misuse of the financial system.[12]

In a nutshell, the legislative proposal presented by the European institutions takes a closer look at the use of technological innovation in the ML and TF field and the proposal of a specific regulation on the subject, named a 'single rulebook', may suggest the need to maximise the harmonisation of the nationals' legislations.[13] Indeed, even though the AML directives that occurred were conceived as self-executing, the parts that were not directly applicable couldn't overcome the fragmentation of the national interpretations of the rules.[14]

One of the key points among the proposed package is the full inclusion of the crypto sector in the regulatory framework; for instance, crypto-assets service providers (CASPs) will be mandated to adhere to due diligence requirements, such as customer verification and suspicious activities' reporting, with specific references on the prohibition of anonymous accounts.

Moreover, the inclusion of crypto asset service providers among the entities subject to AML/CFT rules and the introduction of information requirements for transfers of virtual assets will complement the recent Digital Finance Package, which includes the markets in crypto-assets regulation (known as 'MiCAR'),[15] the regulation of the digital operational resilience (known as 'DORA'),[16] and the DLT pilot regime.[17] In a nutshell, it is possible to notice the growing attention of the European legislator on the distributed ledger technologies. Particularly, the blockchains represent a point of interest because of their own characteristics such as the use of cryptography and the resistance to data tampering.[18]

## 3. The raising and processing of funds: the role of technological innovation

The economic cost of money laundering and terrorism financing cannot be underestimated both in terms of direct and indirect costs. Indirect costs are even greater as they affect the entire economy, potentially impacting financial stability (e.g., as a result of bank runs and loss of foreign investment).

On a large scale, they can cause volatility in international capital flows, undermine good governance, and lead to political instability, eroding trust in governments and institutions.[19]

The impact of criminal activities can be quantified, according to the European Union, as follows: globally, it is estimated that between €715 billion — €1.87 trillion of global GDP (2%-5%) is laundered each year; within the European Union, the 70% of criminal networks employ some form of money laundering to fund their operations and conceal their assets while the 80% of the criminal networks misuse legal business structures for their criminal activities. The EU's financial system and economy saw suspicious activities and transactions for an amount of €117 — €210 billion.[20]

These are examples of the possible misuse of the financial system, which is one aspect that the Standing Committee on AML/CFT, a permanent internal committee within the European Banking Authority, is called upon to oversee. The Standing Committee on Anti-Money Laundering and Countering Terrorist Financing (AMLSC) is composed of 57 representatives; within this group, one competent authority from each Member State has voting rights, while the others are non-voting members. The Committee also includes observers from various organisations, such as the European Insurance and Occupational Pension Authority (EIOPA), the European Securities and Markets Authority (ESMA), the Supervisory Board of the Single Supervisory Mechanism, the European Commission, the EEA EFTA countries represented in the EBA Board of Supervisors, and the EFTA Surveillance Authority.

As mentioned in the paragraphs above, the growing use of technological innovations and infrastructures, such as distributed ledger solutions, is massively used in the ML/FT.

Therefore, it is crucial to investigate on the mechanisms used, not only in relation to traditional systems but also with the awareness that new technologies may significantly impact the topic. One of the specific features of blockchains, such as the transparency of the transactions, allows law enforcement to trace the movement of criminal proceeds from wallet to wallet, making the polar star of the ML/FT investigation, taken from the insight "follow the money" of the anti-mafia judge Giovanni Falcone, a more effective technique than traditional payment methods without transaction records. Criminals have been caught by tracing their crypto transactions to regulated service providers subject to AML regulations. For this reason, the choice to use alternative systems rather than banking and the growth of DeFi have made money laundering easier for criminals.[21] That is, even if money laundering and terrorism financing activities may differ in some points, they often exploit the same vulnerabilities in financial systems and the broader economy that allow for anonymity and opacity in transactions.

As said above, it is crucial to focus on the identification of the sources; preventing the movement of funds and targeting the sources of funding represents the two directions in which to move.

The misuse of financial systems started in traditional finance. For instance, in the case of the 9/11 attacks, the nineteen terrorists opened bank accounts and received several international money transfers. Thereafter, the most stringent regulations on the banking sector, as the responsibilities in carrying out a more detailed Know Your Customer (KYC), brought the financing through the formal banking sector increasing difficulty, as suspicious transactions are more than likely to be flagged up by banks and financial institutions.[22]

If the banking sector is empowering its ML/FT regulation, it is necessary to consider the existence of alternative methods to transfer and process funds. A good example is the hawala, a trust-based money transfer system that can be included in the so-called Informal Value Transfer Systems. The Informal Funds Transfer Systems (hereinafter IFTS), also known as Informal Value Transfer Systems (hereinafter IVTS), represent one of the main methodologies used by criminal organisations and terrorist groups worldwide to transfer money.[23] The IVTS are described as "any system, mechanism, or network of people that receives money for the purpose of making the funds or an equivalent value payable to a third party in another geographic location, whether or not in the same form".[24]

These systems were used for trade financing in circumstances where moving money or other value was considered too dangerous; therefore, the solution was a system able to transfer money without really moving it. In some countries, IVTS-type networks operate in parallel with formal financial institutions or as a substitute or alternative for them, and currently, they are legally used in some countries, even if they must comply with the internal legislative framework; [25] indeed, other countries decided the opposite. In any case, the FAFT recommendations specifically provide the extension of the ML/CF provision on these alternative remittance systems.[26]

The hawala system refers to an informal channel (then representing an IVTS) for transferring funds from one location to another through the so-called hawaladars, regardless of the nature of the transaction and the countries involved. In a nutshell, the hawala works as follows: the sender approaches a hawaladar and asks to transfer a sum of money to a beneficiary in another location. The hawaladar and the sender agree on a password or code that will be used to identify the transaction, and then the transfers are made by the two hawaladars (one at the sender's location and the other at the receiver's location). The two hawaladars will then settle the debit/credit balances by periodic settlements between them. [27]

If compared to blockchain-based technologies for money transfer, it's interesting to note that there are few similarities between the two sharing certain characteristics, such as the decentralisation with no central authority, the convenience of the system, trustworthiness, the fastness of the transactions and specifically if used with criminal aims, anonymity, or pseudonymity. Malicious actors could exploit this fusion for illicit funding, economic destabilisation, regulatory circumvention, or even digital espionage. This is the way some Authors are considering a sort of continuum between the traditional IVTS and their blockchain-based 'new' version.[28]

The use of blockchain in the IVTS is not only a good example of how the technology is used – or misused – as part of the current methodologies to founds handling, but it emphasised how it is frequently conceived as a tool to shape the reality but not even as an instrument capable of creating an autonomous structure.

Quite the opposite, the debate concerning the changing role of the cryptocurrencies represent the pivotal theme. After the advent of bitcoin, the common idea was that it was an alternative to the fiat currency: the digital word against the physical one. After that, and especially after the hypothesis of the releasing of a private money as concomitant with the fiat currency[29] it must be admitted that the panorama is changing. Along with new form of currencies such as stablecoins and the CBDC, the tokenisation of assets has posed a serious concern on new form of 'money' such as the so-called deposit tokens. There are already some case studies that refer to deposit tokens as «the goal is the additional creation of programmable money that could be used within the framework of smart contracts, which in turn permits for more efficient transactions and refined payment controls»[30] and that they will be «become a widely used form of money within the digital asset ecosystem, just as commercial bank money in the form of bank deposits makes up over 90% of circulating money today».[31] Regarding to that, the Bank of International Settlements states that by becoming part of the traditional deposit transfer process, the deposit token should be able to be integrated into the existing regulatory and supervisory framework and, thanks to the the ability to use a unique transaction registry, transactions can be tracked with the required confidentiality through the use of cryptographic techniques but, at the same time, giving important information with regards to the collection of data.

Digital currencies, as the expression of these goods that DLT has made possible, will then become a useful field of application not only to assess the resilience of current legal categories to digital goods, but most of all to identify the opportunities that digital currencies can create, as far as they are of interest here, related to the discipline of AML – a domain that will need further research in order to discuss both the technical and legal implications.

## 4. Technology and finance: a new design for the AML/CFT investigation?

In this scenario, it appears that international cooperation must be one of the key points of the new strategy. Identifying associates in counter-terrorism cases relies on following the money trail, both for known and new actors: the relationship between money laundering, terrorism financing and the financial system is a familiar theme, but the analysis of this topic continues to evolve.[32]

The case of hawala is just an example that shows the consequences of the connection between traditional systems and new DL-based solutions. As criminal activities are changing, the same needs to be done from the other side; the FATF is pushing on the need for a depth promotion of the financial investigations and asset recovery efforts, specifically stating that competent authorities should use and adapt as necessary, traditional law enforcement techniques as well as virtual asset-specific techniques, including the develop, access and training relating to blockchain analytics and monitoring tools.[33] For instance, some Authors proposed a blockchain-enabled transaction scanning method that uses outlier detection and rapid movement funds rules to detect anomalies in bank transaction history[34] as well as as the possibility of the deanonymization of transactions in order to determine the relationships between users and their transactions[35] and how to implement a scheme to detect whether a transaction can be identified as abnormal.[36] The importance to focus on the wider range to possibilities arose from DLTs is because not only do cryptocurrencies represent a new way to raise funds for illicit purposes, but the current fintech innovations (such as the so-called tokenisation) may allow the exchange of other assets to circumvent the current AML/CFT regulations.[37]

Back in 2020, the Commission proposed the creation of a network of counter-terrorism financial investigators to facilitate the exchange of investigative techniques and experience, improving investigators' analysis and understanding of trends and emerging risks. Thus, the Council of the European Union considered prioritising the establishment of EU-level AML/CFT supervision and the coordination and support mechanism for the nationals Financial Intelligence Units, a perspective already considered by the Commission.[38]

That is, in 2021, the proposal for the establishment of an Anti-Money Laundering Authority (hereinafter AMLA) was brought to the table.[39] To validate this decision, a case study recently released by one of the National FIUs shows the potential of blockchain technology for illicit purposes, such as the exchange of virtual assets, as well as for their detection through blockchain analysis software, especially in the context of transnational activities.[40]

Nine cities were selected to host the headquarter of the AMLA (Brussels, Frankfurt, Dublin, Madrid, Paris, Riga, Vilnius, Vienna, and Rome), but Frankfurt happened to be chosen; the AMLA will be based in the city that is already home to the European Central Bank (ECB) and European Insurance and Occupational Pensions Authority (EIOPA).

The provisional agreement reached by the European institutions grants AMLA additional powers to supervise certain credit and financial institutions directly, including crypto asset service providers, if they are deemed high-risk or operate across borders.

As one of the main points of the proposal, the AMLA will select high-risk credit and financial institutions in several member states to be supervised by joint supervisory teams led by the AMLA; the competent teams will carry out all necessary activities to ensure the best possible investigation of suspected ML/FT situations.

In the first selection process, the above-mentioned authority will supervise up to 40 groups and entities. For those entities not selected, AML/CFT supervision will remain primarily the responsibility of the competent national authorities.

Each Member State will ensure that all regulated entities established in its territory are subject to adequate and effective supervision by one or more supervisors with a risk-based approach.

These provisions are not only relevant to the financial sector, as the non-financial sector will also be supervised. The non-financial sector will, therefore, be under the supervision of specialised teams, so-called supervisory colleges; technical standards for their proper functioning will be developed by the AMLA itself.

These provisions have been called after the massive use of new technologies — and blockchains are one of them — demonstrates the importance of an in-depth understanding of their technical characteristics and how they interact with the financial and non-financial domains, including in instances where enhanced anonymity features are involved, and the diversity of different types of virtual assets (FATF Glossary defines a virtual asset as "a digital representation of value that can be digitally traded or transferred and can be used for payment or investment purposes." It also clarifies that "VAs do not include digital representations of fiat currencies, securities, and other financial assets that are already covered elsewhere in the FATF Recommendations) used for criminal purposes.[41]

## 5. Conclusion

The increasing awareness of the use of distributed ledger technology in the financial sector made European institutions bring to the table a new set of legislative proposals. Not only to address the need for consent to the use of these technologies, as in the case of the digital finance package, but also to prevent its misuse. It is a fact that criminals (as money launderers and terrorists) try to gain from the loopholes in the current legislative framework to escape from identification by competent authorities and that new possibilities around new forms of 'money' are a point of interest that need to be addressed by further research.

This was the motivation to propose the new anti-money laundering package, to strengthen the AML/CFT laws, and to enlarge their scope to include new technologies. The analysis must be driven by the understanding of the need for a multidisciplinary approach; the establishment of a new competent authority, named AMLA, and the empowerment of the role of financial intelligence units are the evidence of the undertaken path.

**Acknowledgement**

The author is a student of the national PhD programme in Blockchain & DLT of the University of Camerino (Italy), supported by a grant from the National Recovery and Resiliency Plan (NRRP) and Blockchain Italia srl. The research activities are carried out thanks to the support of BABEL, research unit in the Department of Economics and Management of the University of Florence (Italy).

**Notes and References**

[12] Di Vizio, F., "Prevenzione e investigazioni: l'uso di IA, big data e soluzioni tecnologiche in ambito finanziario e nel contrasto al riciclaggio (AML) e al finanziamento del terrorismo (CFT)" in *DisCrimen*, accessed 11 January 2024 https://discrimen.it/prevenzione-e-investigazioni-luso-di-ia-big-data-e-soluzioni-tecnologiche-in-ambito-finanziario-e-nel-contrasto-del-riciclaggio-aml-e-al-finanziamento-del-terrorismo-cft/

[13] Proposal for a Regulation of the European Parliament and of the Council on the prevention of the use of the financial system for the purposes of money laundering or terrorist financing, COM/2021/420 final in EUR-Lex - 52021PC0420 – EN

[14] Alfieri, G., "Antiriciclaggio: in arrivo il codice unico UE" in *IPSOA Quotidiano*, 3 August 2021, accessed 22 January 2024 www.ipsoa.it/documents/quotidiano/2021/08/03/antiriciclaggio-arrivo-codice-unico-ue

[15] Regulation (EU) 2023/1114 of the European Parliament and of the Council of 31 May 2023 on markets in crypto-assets, and amending Regulations (EU) No 1093/2010 and (EU) No 1095/2010 and Directives 2013/36/EU and (EU) 2019/1937, OJ L 150, 9 June 2023

[16] Regulation (EU) 2022/2554 of the European Parliament and of the Council of 14 December 2022 on digital operational resilience for the financial sector and amending Regulations (EC) No 1060/2009, (EU) No 648/2012, (EU) No 600/2014, (EU) No 909/2014 and (EU) 2016/1011, OJ L 333, 27 December 2022

[17] Regulation (EU) 2022/858 of the European Parliament and of the Council of 30 May 2022 on a pilot regime for market infrastructures based on distributed ledger technology, and amending Regulations (EU) No 600/2014 and (EU) No 909/2014 and Directive 2014/65/EU OJ L 151, 2 June 2022

[18] Schär F., Berentsen A., "Bitcoin, Blockchain, and Cryptoassets, a Comprehensive Introduction" Cambridge (MA) London, *MIT Press*, 2020; Abhishek T., Benarji C., "Recalibrating the Banking Sector with Blockchain Technology for Effective Anti-Money Laundering Compliances by Banks" *Sustainable Futures*, 2023, https://doi.org/10.1016/j.sftr.2023.100107; Sai BDS and others, "A Decentralised KYC Based Approach for Microfinance Using Blockchain Technology" (2023) 1 *Cyber Security and Applications*, https://doi.org/10.1016/j.csa.2022.100009.

[19] International Monetary Fund, *Financial Crimes Hurt Economies and Must Be Better Understood and Curbed* (IMF, 12 July 2023) accessed 05 February 2024 www.imf.org/en/Blogs/Articles/2023/12/07/financial-crimes-hurt-economies-and-must-be-better-understood-and-curbed

[20] European Council, *Infographic: what is money laundering* in www.consilium.europa.eu/en/infographics/anti-money-laundering/.

[21] Benson, V., Turksen, U., Adamyk, B., "Dark Side of Decentralised Finance: A Call for Enhanced AML Regulation Based on Use Cases of Illicit Activities" *Journal of Financial Regulation and Complianc*e, 32(2023), accessed 29 January 2024 https://doi.org/10.1108/JFRC-04-2023-0065

[22] Salami, I., "Terrorism Financing with Virtual Currencies: Can Regulatory Technology Solutions Combat This?", *Studies in Conflict & Terrorism* 41(2018) accessed 29 January 2024 https://doi.org/10.1080/1057610X.2017.1365464

[23] Barberini, D., "Informal Funds Transfer Systems. I sistemi di trasferimento informale di denaro", *Rivista italiana dell'antiriciclaggio* 4 219-229 (2022)

[24] US Department of the Treasury Financial Crimes Enforcement Network, "Informal Value Transfer Systems" *FinCen Advisory*, 33 (2003) accessed 06 February 2024 www.fincen.gov/sites/default/files/advisory/advis33.pdf

[40] Unità di Informazione Finanziaria per l'Italia, "Casistiche di riciclaggio e di finanziamento del terrorismo" *Quaderni dell'antiriciclaggio* 21 (2023) accessed 30 January 2024 https://uif.bancaditalia.it/pubblicazioni/quaderni/2023/quaderno-21-2023/quaderno-21-2023.pdf

[41] International Monetary Fund "Virtual Assets and Anti-Money Laundering and Combating the Financing of Terrorism (1) Some Legal and Practical Considerations", 14 October 2021, accessed 28 January 2024 www.imf.org/en/Publications/fintech-notes/Issues/2021/10/14/Virtual-Assets-and-Anti-Money-Laundering-and-Combating-the-Financing-of-Terrorism-1-463654.